\documentclass[12pt]{article}

\usepackage{amsmath,amssymb}
\usepackage{slashed}
\usepackage{xcolor} 
\usepackage{graphicx}
\usepackage{url}
\usepackage{cite}

\newcommand{\eref}[1]{Eq.~\eqref{eq:#1}}

\newcommand{\sref}[1]{Section~\ref{sec:#1}}
\newcommand{\cref}[1]{Chapter~\ref{ch:.#1}}
\newcommand{\tref}[1]{Table~\ref{tab:#1}}

\newcommand{\nnl}{\nonumber \\}

\newcommand{\beq}{\begin{equation}} 
\newcommand{\eeq}{\end{equation}} 
\newcommand{\ba}{\begin{array}}  
\newcommand{\ea}{\end{array}} 
\newcommand{\bea}{\begin{eqnarray}}  
\newcommand{\eea}{\end{eqnarray} }  
\newcommand{\be}{\begin{eqnarray}}  
\newcommand{\ee}{\end{eqnarray} }  
\newcommand{\bal}{\begin{align}}
\newcommand{\eal}{\end{align}}   
\newcommand{\bi}{\begin{itemize}}  
\newcommand{\ei}{\end{itemize}}  
\newcommand{\ben}{\begin{enumerate}}  
\newcommand{\een}{\end{enumerate}}  
\newcommand{\bc}{\begin{center}}
\newcommand{\ec}{\end{center}} 
\newcommand{\bt}{\begin{table}}
\newcommand{\et}{\end{table}}  
\newcommand{\btb}{\begin{tabular}}
\newcommand{\etb}{\end{tabular}}  
\newcommand{\bvec}{\left ( \ba{c}}
\newcommand{\evec}{\ea \right )}

\newcommand{\cO}{{\mathcal O}} 
 
\newcommand{\cL}{{\mathcal L}}

\def\hc{{\rm h.c.}} 

\newcommand{\eps}{\epsilon}

\begin{document}

\vspace*{-2cm}

\begin{center}
\vspace*{15mm}

\vspace{1cm}
{\large \bf 
Effective field theory approach to LHC Higgs data
} \\
\vspace{1.4cm}

Adam Falkowski

 \vspace*{.5cm} 
Laboratoire de Physique Th\'{e}orique, Bat.~210, Universit\'{e} Paris-Sud, 91405 Orsay, France.

\vspace*{.2cm} 

\end{center}

\vspace*{10mm} 
\begin{abstract}\noindent\normalsize
I review the effective field theory approach to LHC Higgs data.


\end{abstract}

\vspace*{3mm}

\vspace{12cm}

{\it Invited  review prepared for Pramana - Journal of Physics}

\newpage

\renewcommand{\theequation}{\arabic{section}.\arabic{equation}} 

\section{Introduction}
\setcounter{equation}{0} 

The standard model (SM) of particle physics was proposed back in the 60s as a theory of quarks and leptons interacting via strong, weak, and electromagnetic forces \cite{Weinberg:1967tq}. 
It is build on the following principles 
\ben
\item[\#1] The basic framework is that of a relativistic quantum field theory,
with interactions between particles described by a local Lagrangian. 
\item[\#2] The Lagrangian is invariant under the linearly realized local $SU(3) \times SU(2) \times U(1)$ symmetry. 
\item[\#3] The vacuum state of the theory preserves only  $SU(3)  \times U(1)$ local symmetry, as a result of the Brout-Englert-Higgs mechanism \cite{Englert:1964et,Higgs:1964ia,Guralnik:1964eu}. 
The spontaneous breaking of the $SU(2) \times U(1)$ symmetry down to $U(1)$ arises due to a vacuum expectation value  (VEV) of a scalar field transforming as $(1,2)_{1/2}$ under the local symmetry. 
\item[\#4] Interactions are {\em renormalizable}, which means that only interactions up to the canonical  mass dimension 4 are allowed in the Lagrangian. 
\een 
Given the experimentally observed matter content (3 families of quarks and leptons), these rules completely specify the theory up to 19 free parameters. 
The local symmetry implies the presence of spin-1 vector bosons which mediate the strong and electroweak forces. 
The breaking pattern of the local symmetry ensures that the carriers of the strong and electromagnetic force are massless, whereas tho carries of the weak force are massive. 
Finally,  the particular realization of the Brout-Englert-Higgs  mechanism in the SM leads to the emergence of exactly one spin-0 scalar boson - the famous Higgs boson \cite{Higgs:1964pj,Higgs:1966ev,Kibble:1967sv}.  

The SM  passed an incredible number of experimental tests. 
It correctly describes the rates and differential distributions of particles produced  in high-energy collisions; 
a robust deviation from the SM predictions has never been observed. 
It predicts very accurately many properties of elementary particles, such as the magnetic and electric dipole moments,  as well as certain properties of simple enough composite particles, such as atomic energy levels. 
The discovery  of a 125 GeV boson at the Large Hadron Collider (LHC)~\cite{Aad:2012tfa,Chatrchyan:2012ufa} nails down the last propagating degree of freedom predicted by the SM. 
Measurements of its production and decay rates  vindicates the simplest realization of the Brout-Englert-Higgs mechanism, in which a VEV of a  {\em single} SU(2) doublet field spontaneously breaks the electroweak  symmetry.
Last not least, the SM is a consistent quantum theory, whose validity range extends to energies all the way up to the Planck scale (at which point the gravitational interactions become strong and can no longer be neglected).   

Yet we know that the SM is not the ultimate theory. 
It cannot account for dark matter, neutrino masses, matter/anti-matter asymmetry,  and cosmic inflation, which are all experimental facts. 
In addition, some theoretical or esthetic arguments (the strong CP problem, flavor hierarchies, unification, the naturalness problem) suggest that the SM should be extended. 
This justifies the ongoing searches for {\em new physics}, that is particles or interactions not predicted by the SM. 

In spite of good arguments for the existence of new physics, a growing body of evidence suggests that, at least up to energies of a few hundred GeV, the fundamental degrees of freedom are those of the SM. 
Given the absence of any direct or indirect collider signal of new physics, it is reasonable to  assume that new particles from beyond the SM  are much heavier than the SM particles.
If that is correct, physics at the weak scale can be adequately described using {\em effective field theory} (EFT) methods. 

In the EFT framework adopted here the assumptions $ \#1 \dots \#3$ above continue to be valid.\footnote{%
One could consider a more general EFT where the assumptions $\#2$ and $\#3$ are also relaxed and the electroweak symmetry is realized {\em non-linearly} \cite{Grinstein:2007iv,Alonso:2012px,Isidori:2013cla,Buchalla:2013rka}.
In that case, the Higgs boson is introduced as a singlet of the local symmetry, rather than as a part of an $SU(2)$ doublet.
} 
Thus, much as in the SM, the Lagrangian is constructed from gauge invariant operators  involving the SM fermion, gauge, and Higgs fields. 
The difference is that the assumption $\#4$ is dropped and interactions with arbitrary large mass dimension $D$ are allowed.  
These interactions can be organized in a  systematic expansion in $D$. 
The leading order term in this expansion is the SM Lagrangian with operators up to $D= 4$. 
All  possible effects of heavy new physics are encoded in operators with $D>4$,  which are suppressed in the Lagrangian by appropriate powers of 
the mass scale $\Lambda$.    
Since all $D=5$ operators violate lepton number and are thus stringently  constrained by experiment, the leading corrections to the Higgs observables are expected from $D=6$ operators suppressed by $\Lambda^2$ \cite{Buchmuller:1985jz}. 
I will assume that the operators with $D>6$ can be ignored, which is always true for $v \ll \Lambda$.  

This review discusses the interpretation of the LHC data on the Higgs boson production and decay in the framework of an EFT beyond the SM. 
For practical reasons, three more assumptions about higher-dimensional operators are adopted: 
\bi
\item The  baryon and lepton numbers are conserved.  
\item The coefficients of operators involving fermions are flavor conserving and universal, except for Yukawa-type operators, which are aligned with the corresponding SM Yukawa matrices.\footnote{This assumption is largely practical, because there is little experimental information about Higgs couplings to the 1st and 2nd generation fermions.  
Currently, these couplings are probed indirectly \cite{Goertz:2014qia,Altmannshofer:2015qra}, while in the future some may be probed directly via exclusive Higgs decays to a photon and a meson \cite{Bodwin:2013gca,Kagan:2014ila}.  
 } 
\item The corrections from $D=6$ operators to the Higgs signal strength are subleading compared to the SM contribution. 
\ei 
Other than that, the discussion will be model-independent. 


In the following section I review the SM Lagrangian, in order to prepare the ground and fix the notation. 
The part of the $D=6$ effective Lagrangian relevant for Higgs studies is discussed in  \sref{effl}. 
The dependence of the Higgs signal strength measured at the LHC on the effective Lagrangian parameters  is summarized in \sref{obs}. 
The experimental results and the current model-independent constraints on the $D=6$ parameters are discussed in \sref{constraints}.  
The bibliography contains  a number of references where an EFT-inspired approach to physics of the 125~GeV Higgs at the LHC is exercised; 
citation complaints are welcome for omitted papers belonging to that category.

\section{Standard Model Lagrangian}
\setcounter{equation}{0} 
\label{sec:sm}

I  start by summarizing the SM Lagrangian and defining my notation.  

The SM Lagrangian is invariant under the global Poincar\'e symmetry (Lorentz symmetry + translations), 
and a local symmetry with the gauge group  $G_{\rm SM } = SU(3)_C \times SU(2)_L \times U(1)_Y$. 
The fields building the SM Lagrangian fill representations of these symmetries.  
The field content of the SM is the following: 
\bi
\item Vector fields $ G_\mu^a$, $W_\mu^i$, $B_\mu$, where  $i=1\dots 3$ and $a = 1 \dots 8$.  
They transform as four-vectors under the Lorentz symmetry and are the gauge fields of the  $G_{\rm SM }$ group. 
\item 3 generations of fermionic fields $q = (u, V_{\rm CKM}  d)$, $u^c$, $d^c$, $\ell  = (\nu,e)$, $e^c$. 
They transform as 2-component spinors under the Lorentz symmetry.\footnote{%
Throughout this review I use the 2-component spinor notation for fermions; in all instances I follow the conventions of Ref.~\cite{Dreiner:2008tw}.}
The transformation properties under  $G_{\rm SM }$ are listed in \tref{SM_rep}. 
\item Scalar field $H  = (H^+,H^0)$ transforming  as $(1,2)_{1/2}$ under $G_{\rm SM}$. 
I also define $\tilde H_i = \eps_{ij} H^*_j$ that transforms as $(1,2)_{-1/2}$. 
\ei   

\begin{table}
\bc
\begin{tabular}{c|c|c|c|c}
   & $SU(3)_C$ & $SU(2)_L$ & $U(1)_Y$  
   \\  \hline 
$q = \bvec u \\ d \evec$ & {\bf 3} & {\bf 2} & 1/6 
\\  \hline 
$u^c$ & ${\bf \bar 3}$ & {\bf 1} & -2/3   
\\  \hline 
$d^c$ & ${\bf \bar 3}$ & {\bf 1} & 1/3 
\\  \hline 
$\ell = \bvec \nu \\ e \evec$ &  {\bf 1} & {\bf 2} & -1/2 
\\  \hline 
$e^c$     &  {\bf 1} & {\bf 1} & 1  
\\  \hline 
$H$     &  {\bf 1} & {\bf 2} & 1/2  
\end{tabular}
\ec
\caption{
\label{tab:SM_rep}
Representation of the SM scalar and fermion fields under the SM gauge group.}
\end{table}

The SM Lagrangian can be split as 
\beq
\label{eq:lsm}
\cL^{\rm SM} =  \cL^{\rm SM}_{\rm V}  +   \cL^{\rm SM}_{\rm F}  +   \cL^{\rm SM}_{\rm H}+    \cL^{\rm SM}_{\rm Y}.    
\eeq 
The first term above contains gauge invariant kinetic terms for the vector fields: 
\beq 
\label{eq:lsmv}
\cL^{\rm SM}_{\rm V} = -{1 \over 4 g_s^2} G_{\mu \nu}^a G_{\mu \nu}^a  - {1 \over 4 g_L^2} W_{\mu \nu}^i W_{\mu \nu}^i  -  {1 \over  4 g_Y^2}    B_{\mu \nu} B_{\mu \nu},   
\eeq 
 where   $g_s$, $g_L$, $g_Y$ are gauge couplings  of $SU(3)_C \times SU(2)_L \times U(1)_Y$, here  defined as the normalization of the appropriate gauge kinetic term. 
I also define the electromagnetic coupling $e = g_L g_Y/\sqrt{g_L^2 + g_Y^2}$, and the Weinberg angle $s_\theta = g_Y/\sqrt{g_L^2 + g_Y^2}$.  
The field strength tensors are given by 
\beq
B_{\mu \nu} = \partial_\mu B_\nu - \partial_\nu B_\mu,  
\  \ 
W_{\mu \nu}^i = \partial_\mu W_\nu^i - \partial_\nu W_\mu^i  +   \eps^{ijk}  W_\mu^j W_\nu^k,  
\  \
G_{\mu \nu}^a = \partial_\mu G_\nu^a - \partial_\nu G_\mu^a  +    f^{abc}G_\mu^b G_\nu^c.   
\eeq
where $\eps^{ijk}$ and $f^{abc}$  are the totally anti-symmetric structure tensors of $SU(2)$ and $SU(3)$.  

The second term in \eref{lsm} contains covariant kinetic terms of the fermion fields: 
\beq
\label{eq:lsmf}
\cL^{\rm SM}_{\rm F}  = i\bar q \bar \sigma_\mu D_\mu q     +  i u^c \sigma_\mu  D_\mu \bar u^c  +   i d^c \sigma_\mu  D_\mu \bar d^c +   i \bar \ell \bar \sigma_\mu D_\mu \ell   +  i e^c \sigma_\mu  D_\mu \bar e^c.
\eeq 
Each fermion field is a 3-component vector in the generation space. 
I assume all the rotations needed to put fermions in the mass eigenstate basis have already been made; in the SM the only residue of these rotations is the CKM matrix appearing in the definition of the quark doublet components. 
The covariant derivatives are  defined as 
\beq 
 D_\mu f =   \left (\partial_\mu  - i  G_\mu^a   T^a_f -   i   W_\mu^i T^i_f   -  i Y_f  B_\mu \right ) f .
 \eeq 
Here $T^a_f = (\lambda^a,-\lambda^a,0)$ for $f$ in the triplet/anti-triplet/singlet representation of SU(3), where $\lambda^a$ are  Gell-Mann matrices;  
$T^i_f = (\sigma^i/2,0)$ for $f$ in the doublet/singlet representation of SU(2); 
$Y_f$ is the U(1) hypercharge. 
The electric charge is given by $Q_f = T^3_f  + Y_f$. 
 
The third term in  \eref{lsm} contains Yukawa interactions between the Higgs field and the fermions:  
\beq
\label{eq:lsmy}
\cL^{\rm SM}_{\rm Y} =  -  \tilde H^\dagger   u^c y_u q  -  H^\dagger d^c  y_d q -  H^\dagger e^c   y_e \ell     + \hc,  
\eeq 
where $y_f$ are $3 \times 3$ diagonal matrices. 

The last term in \eref{lsm} are the Higgs kinetic and potential terms: 
\beq
\label{eq:lsmh}
\cL^{\rm SM}_{\rm H} = D_\mu H^\dagger D_\mu H  + \mu_H^2 H^\dagger H - \lambda (H^\dagger H)^2 , 
\eeq  
where the covariant derivative acting on the Higgs field is 
\beq
D_\mu H = \left ( \partial_\mu  -  {i \over 2} W_\mu^i   \sigma^i  -  {i  \over 2} B_\mu  \right ) H . 
\eeq


Because of the negative mass squared term $\mu_H^2$ in the Higgs potential the Higgs  field gets a VEV, 
\beq
\langle H \rangle = {1 \over \sqrt 2} \bvec 0 \\ v \evec,  \qquad \mu_H^2 = \lambda v^2 . 
\eeq  
This generates mass terms for $W_\mu^i$ and $B_\mu$ and a field rotation is needed to diagonalize the mass matrix. 
The mass eigenstates are defined to the electroweak vector fields by 
\bea 
W_\mu^1  = {g_L \over \sqrt 2 } \left ( W_\mu^+ + W_\mu^- \right ), 
&\qquad&  
W_\mu^3 = {g_L \over \sqrt{g_L^2 + g_Y^2} } \left ( g_L Z_\mu + g_Y A_\mu \right ), 
\nnl
W_\mu^2  = {i g_L\over \sqrt 2 } \left ( W_\mu^+ - W_\mu^- \right ), 
& \qquad  &
B_\mu =  {g_Y \over \sqrt{g_L^2 + g_Y^2} } \left ( - g_Y Z_\mu + g_L A_\mu \right ). 
\eea
With this definition, the mass eigenstates  such that their quadratic terms are canonically normalized and their mass terms are diagonal: 
\beq
\label{eq:SM_lvkin}
\cL_{V, \rm kin}^{\rm SM} = -{1 \over 2} W_{\mu \nu}^+  W_{\mu \nu}^- -   {1 \over 4} Z_{\mu \nu} Z_{\mu \nu} 
-   {1 \over 4} A_{\mu \nu} A_{\mu \nu}  + m_W^2  W_{\mu}^+  W_{\mu}^-  + {m_Z^2  \over 2 } Z_\mu Z_\mu,
\eeq  
where the  W and Z boson masses are 
\beq
m_W = {g_L v \over 2}, \qquad 
m_Z = {\sqrt{g_L^2 + g_Y^2} v\over 2}. 
\eeq
The SM fermions (except for the neutrinos) also acquire masses  after electroweak symmetry breaking via the Yukawa interactions in \eref{lsmy}. 
I choose  a basis in the fermion flavor space where the Yukawa interactions are diagonal, in which case the fermion masses are given by 
\beq
m_{f_i} = {v \over \sqrt 2} [y_f]_{ii}. 
\eeq 
Interactions of the gauge boson mass eigenstates with fermions are given by 
\bea 
\label{eq:SM_vff}
\cL^{\rm SM }_{v ff} & =  &
 e A_\mu \sum_{f  \in u,d,e} Q_f (\bar f \bar \sigma_\mu f  + f^c \sigma_\mu  \bar f^c)   + g_s G_\mu^a \sum_{f  \in u,d} (\bar f \bar \sigma_\mu T^a f  + f^c \sigma_\mu T^a \bar f^c) 
\nnl &+& 
{g_L \over \sqrt 2}  \left ( 
W_\mu^+ \bar u \bar \sigma_\mu V_{\rm CKM} d  
+  W_\mu^+  \bar \nu \bar \sigma_\mu e+  \hc \right )
\nnl &+& \sqrt{g_L^2 + g_Y^2} Z_\mu      \left [  
   \sum_{f \in u, d,e,\nu} \bar f \bar \sigma_\mu (  T^3_f  -   s^2_\theta Q_f) f  
+   \sum_{f \in u, d,e} f^c \sigma_\mu (-  s^2_\theta Q_f )\bar  f^c  \right ]. 
\nnl 
\eea

Finally, I move to the Higgs  sector. 
After electroweak symmetry breaking,  the  Higgs doublet field can be conveniently  written as 
\beq
H = {1 \over \sqrt 2} \bvec \sqrt{2} G^+ \\ v + h + i G^0 \evec.  
\eeq 
The fields $G^0$ and $G^+$ do not correspond to new physical degrees of freedom (they kinetically mix with the massive gauge bosons and can be gauged away). 
From now on,  I will work in the unitary gauge and  set $G^\pm = 0 = G^0$. 
The star of this review - the scalar field $h$ - is called {\em the Higgs boson}. 
Its mass  can be  expressed by the parameters of the Higgs potential  as 
\beq
m_h^2 = 2 \mu_H^2 = 2 \lambda v^2.  
\eeq 
The interactions in the SM Lagrangian involving a single Higgs boson  are the following   
\beq
\label{eq:SMh}
\cL^{\rm SM}_h = 
 {h \over v}   \left [ {g_L^2 v^2 \over 2} W_\mu^+ W_\mu^- +   {(g_L^2 + g_Y^2) v^2 \over 4}    Z_\mu Z_\mu \right ] 
 -  {h \over v}  \sum_f  m_f \left ( f f^c + \hc  \right ).
\eeq 
Roughly speaking,  the  Higgs boson couples to mass, in the sense that it couples to pairs of SM particles with the strength proportional to their masses (for fermions) or masses squared (for bosons).    
Since all the masses have been measured  by experiment, 
the strength of Higgs boson interactions in the SM is precisely predicted and contains no free parameters.  

I conclude this section  with a  summary of the SM parameters used in this review. 
For the Higgs boson mass I take $m_h = 125.09$~GeV, which is the central value of the recent ATLAS and CMS combination of  mass measurements  \cite{Aad:2015zhl}. 
The gauge boson masses are  $m_W = 80.385$~GeV \cite{Group:2012gb} , and $m_Z = 91.1875$~GeV \cite{ALEPH:2005ab}. 
The Higgs VEV is calculated at from the muon lifetime 
 (equivalently, from the Fermi constant $G_F = 1/\sqrt 2 v^2 = 1.16637 \times 10^{-5}~{\rm GeV}^{-2}$~\cite{Beringer:1900zz}), corresponding to  $v = 246.221$~GeV.
The electroweak couplings at the Z boson mass scale  are extracted from $m_Z$ and  the  electromagnetic structure constant $\alpha(m_Z) = 7.755 \times 10^{-3}$ \cite{Burkhardt:2011ur}, and the strong coupling from $\alpha_s(m_Z) = 1.172 \times 10^{-3}$~\cite{Beringer:1900zz}. 
To evaluate corrections to the  Higgs observables I will use the couplings run up to the scale $m_h$: $g_s = 1.187$, $g_L = 0.643$, $g_Y =  0.358$. 
The light fermion masses are also evaluated at the scale $m_h$: the relevant ones are  $m_b = 2.76$~GeV, $m_\tau = 1.78$~GeV, and $m_c = 0.62$~GeV. 
For the top mass I take $m_t = 173.2$~GeV.

\section{Dimension Six Lagrangian}
\setcounter{equation}{0} 
\label{sec:effl}

We consider the effective Lagrangian of the form, 
\beq
\label{eq:leff}
\cL_{\rm eff}  = {\cal L}^ {\rm SM}  +   {\cal L}^ {D=6}, \qquad   {\cal L}^ {D=6} = {1 \over v^2} \sum_\alpha c_\alpha O_\alpha,  
\eeq 
where $\cL^{\rm SM}$ is the SM Lagrangian discussed in \sref{sm},  and $O_\alpha$ is a complete basis of $SU(3) \times SU(2) \times U(1)$ invariant $D=6$ operators constructed out of the SM fields.  
In general, such a basis contains 2499 independent operators after imposing baryon and lepton number conservation \cite{Alonso:2013hga}. 
One of the assumptions in this review is  that coefficients of $D=6$ operators are flavor universal, which brings the number of independent parameters  down to 76.   
Furthermore, only 9 combinations of these operators  will be relevant for a completely general description of the Higgs signal strength measurements considered later in this review.  

One can choose a complete, non-redundant basis of operators in many distinct (though equivalent) ways.   
Here we work with the so-called {\em Higgs basis} introduced in Ref.~\cite{HXSWGbasis} and inspired by Refs.~\cite{Gupta:2014rxa,Pomarol:2014dya}.\footnote{%
Other popular choices in the Higgs-related literature are the Warsaw basis~\cite{Grzadkowski:2010es,Alonso:2013hga}, and the SILH basis \cite{Giudice:2007fh,Contino:2013kra}.}
The basis is spanned by  particular combinations of $D=6$ operators.
Each of these combinations maps to an interaction term of the SM mass-eigenstates in the tree-level  effective Lagrangian. 
The coefficients multiplying these combinations in the Lagrangian are called the {\em independent couplings}.  
The single Higgs couplings to pairs of gauge bosons and fermions are chosen among the independent couplings.
The advantage of this basis is that the independent couplings are related in a simple way to observables in Higgs physics.   

Most often, an $SU(3) \times SU(2) \times U(1)$ invariant  operator gives rise to more than one interaction term of mass eigenstates. 
This leads to relations between various couplings in the effective Lagrangian. 
Therefore,  several of these couplings are not free but can be expressed in terms of the independent couplings; they are called the {\em dependent} couplings.
For example, at the level of the $D=6$ Lagrangian, the W boson couplings to fermions are dependent couplings,  as they can be expressed in terms of the Z boson couplings to fermions.  
Of course, the choice which couplings are chosen as independent and which are dependent is subjective and dictated by convenience. 

Below I review the part of $D=6$ Lagrangian in the Higgs basis that is  relevant for LHC Higgs observables; see  Ref.~\cite{HXSWGbasis} for the full set of independent couplings and the algorithm to construct the complete $D=6$ Lagrangian. 
In this formalism, by construction, all kinetic terms are canonically normalized, there is  no kinetic mixing between the Z boson and the photon, and there is no correction to the Z boson mass term. 
While, in general,  $D=6$ operators  do generate mixing and mass corrections, the canonical form can always be recovered by using equations of motion, integration by parts, and redefinition of fields and  couplings. 
Thus, the kinetic and mass terms for the electroweak  gauge bosons  are those in \eref{SM_lvkin},  except for  the correction to the W boson mass term: 
 $\Delta \cL_{\rm kinetic}^{D=6} =  2 \delta m {g^2 v^2 \over 4} W_\mu^+ W_\mu^-$. 
 The  independent coupling $\delta m$ is a free parameter from the EFT point of view, however it is very well constrained by experiment: 
 $\delta m = (2.6 \pm 1.9)\cdot 10^{-4}$ \cite{Efrati:2015eaa}. 
 Given the precision of LHC data, effects proportional to $\delta m$ are currently not relevant for Higgs searches and will be ignored.   
 
We move to interactions of a single Higgs boson with pairs of SM gauge bosons and fermions. 
The SM interactions of this type were given in  \eref{SMh} and they contain no free parameters. 
Dimension six operators lead to shifts of the couplings in  \eref{SMh}, as well as to the appearance of 2-derivative Higgs  couplings to gauge bosons. 
In the Higgs basis, these effects are parametrized by the following independent couplings: 
\bea 
\label{eq:ind} 
&  \delta c_z, \ c_{zz}, \   c_{z \Box}, \ \ c_{\gamma \gamma}, \ c_{z \gamma},  \  c_{gg}, \  \tilde c_{gg}, \ \tilde c_{zz},  \   \tilde c_{\gamma \gamma}, \  \tilde c_{z \gamma},
 \nnl
& \delta y^u, \   \delta y^d,  \ \delta y^e, \   \sin \phi^u, \   \sin \phi^d,  \  \sin \phi^\ell. 
\eea  
The  couplings in the first line are defined via the Higgs boson couplings  to gauge bosons:  
\bea
\label{eq:hvv}
\Delta \cL_{\rm hvv}^{D=6}   &= & {h \over v} \left [ 
2 \delta c_w   m_W^2 W_\mu^+ W_\mu^- +   \delta c_z   m_Z^2 Z_\mu Z_\mu
\right . \nnl & & \left . 
+ c_{ww}  {g_L^2 \over  2} W_{\mu \nu}^+  W_{\mu\nu}^-  + \tilde c_{ww}  {g_L^2 \over  2} W_{\mu \nu}^+   \tilde W_{\mu\nu}^-  
+ c_{w \Box} g_L^2 \left (W_\mu^- \partial_\nu W_{\mu \nu}^+ + \hc \right )  
\right . \nnl & & \left . 
+  c_{gg} {g_s^2 \over 4 } G_{\mu \nu}^a G_{\mu \nu}^a   + c_{\gamma \gamma} {e^2 \over 4} A_{\mu \nu} A_{\mu \nu} 
+ c_{z \gamma} {e g_L  \over  2 c_\theta} Z_{\mu \nu} A_{\mu\nu} + c_{zz} {g_L^2 \over  4 c_\theta^2 } Z_{\mu \nu} Z_{\mu\nu}
\right . \nnl & & \left .
+ c_{z \Box} g_L^2 Z_\mu \partial_\nu Z_{\mu \nu} + c_{\gamma \Box} g_L g_Y Z_\mu \partial_\nu A_{\mu \nu}
\right . \nnl & & \left . 
+  \tilde c_{gg} {g_s^2 \over 4} G_{\mu \nu}^a \tilde G_{\mu \nu}^a  
+ \tilde c_{\gamma \gamma} {e^2 \over 4} A_{\mu \nu} \tilde A_{\mu \nu} 
+ \tilde c_{z \gamma} {e g_L \over  2 c_\theta} Z_{\mu \nu} \tilde A_{\mu\nu}
+ \tilde c_{zz}  {g_L^2  \over  4 c_\theta^2} Z_{\mu \nu} \tilde Z_{\mu\nu}
\right ], 
\nnl
 \eea 
where the dependent couplings  $\delta c_w$, $c_{ww}$,  $\tilde c_{ww}$, $c_{w \Box}$, and $c_{\gamma \Box}$  can be expressed by the independent couplings as
\bea
\label{eq:np2}
\delta  c_{w} &=&  \delta c_z + 4 \delta m , 
\nnl 
c_{ww} &=&  c_{zz} + 2 s_\theta^2 c_{z \gamma} + s_\theta^4 c_{\gamma \gamma}, 
\nnl 
\tilde c_{ww} & = &  \tilde c_{zz} + 2 s_\theta^2  \tilde c_{z \gamma} + s_\theta^4  \tilde c_{\gamma \gamma} , 
\nnl
c_{w \Box}  &= & {1 \over g_L^2 - g_Y^2} \left [ 
g_L^2 c_{z \Box} + g_Y^2 c_{zz}  - e^2 s_{\theta}^2   c_{\gamma \gamma}  -(g_L^2 - g_Y^2) s_\theta^2  c_{z \gamma} 
\right ],  
\nnl 
  c_{\gamma \Box}  &= &  
  {1 \over g_L^2 - g_Y^2} \left [ 
2 g_L^2 c_{z \Box} + (g_L^2+ g_Y^2) c_{zz}  - e^2  c_{\gamma \gamma}  -(g_L^2 - g_Y^2)   c_{z \gamma} 
\right ].  
  \eea 
The coupling in the second line of \eref{ind}  are  defined via the  Higgs boson couplings to fermions: 
\beq
\label{eq:hff}
\Delta \cL_{\rm hff}^{D=6}  =  - {h \over v} \sum_{f \in u,d,e}    \delta y_f \, e^{i \phi_f }  \, m_f  f^c f  + \hc . 
\eeq
Following my assumption of flavor universal coefficients of dimension-6 operators,  each $\delta y_f$ and  $\phi_f$ is a real number. 
Moreover, the couplings in \eref{hff} are diagonal in the generation space, therefore flavor violating Higgs decays are absent (see Refs.~\cite{Blankenburg:2012ex,Harnik:2012pb} for a discussion of such decays in the EFT language).  

The complete Higgs interaction Lagrangian relevant for this review is given by $\cL_{h}^{\rm SM} +\cL_{vff}^{\rm SM} + \Delta \cL_{\rm hvv}^{D=6} + \Delta \cL_{\rm hff}^{D=6}$ and is parametrized  by the independent couplings in \eref{ind}.
The effect of these couplings  on the LHC Higgs observables will be discussed in the following sections. 
But before that, a comment is in order on other effects of $D=6$ operators  that could, a priori, be relevant. 
First,  in the Higgs basis there are corrections to the Z and W boson interactions in \eref{SM_vff}, parametrized by vertex corrections $\delta g$.   
These would feed indirectly into Higgs observables, such as, for example,  the vector boson fusion (VBF) production cross section or  the $h \to VV^* \to 4$~fermions decays.   
However, there are model-independent constraints on these vertex corrections \cite{Efrati:2015eaa} which ensure that their effects on Higgs observables are too small to be currently observable. 
For this reason I will ignore the vertex corrections in this review. 
Next, $D=6$ operators may induce two classes  of Higgs boson interactions that could affect $h \to VV^* \to 4$~fermions  decays. 
One class is the $h V f f$ vertex-like contact interactions:  
\bea
\label{eq:hvff}
\Delta \cL_{hvff}^{D=6}&=&  \sqrt 2  g_L {h \over v}   W_\mu^+   \left (
 \delta g^{W \ell }_L \bar \nu \bar \sigma_\mu  e
 +   \delta g^{Wq}_L \bar u \bar \sigma_\mu  d
+  \delta g^{Wq}_R  u^c  \sigma_\mu   \bar d^c
 + \hc  \right )
\nnl 
&+& 2 {h \over v}  \sqrt{g_L^2 + g_Y^2} Z_\mu 
\left [ \sum_{f = u,d,e,\nu}    \delta g^{Zf}_L  \bar f \bar \sigma_\mu f  +  
\sum_{f = u,d,e}   \delta g^{Zf}_R  f^c \sigma_\mu \bar f^c  \right ] . 
\eea
In the Higgs basis, the parameters $\delta g$ above are equal to the corresponding vertex corrections to the SM couplings in \eref{SM_vff}. 
Given the constraints on the $\delta g$'s in Ref.~ \cite{Efrati:2015eaa}, the LHC Higgs studies cannot be currently sensitive to the vertex-like Higgs interactions, therefore they are neglected in this analysis. 
The other  class  is the dipole-like contact interactions:  
\bea
\Delta \cL_{hdvff}^{D=6}  &=& 
- {h \over 4 v^2 }  \left [ 
g_s \sum_{f \in u,d}   d_{Gf}  f^c  \bar \sigma_{\mu \nu} T^a    f   G_{\mu\nu}^a 
+ e  \sum_{f \in u,d,e}  d_{Af}   f^c  \bar  \sigma_{\mu \nu}   f   A_{\mu\nu} 
\right . \nnl && \left . 
+ \sqrt{g_L^2 + g_Y^2}  \sum_{f \in u,d,e}  d_{Zf}  f^c  \bar  \sigma_{\mu \nu}   f   Z_{\mu\nu}
+ \sqrt {2} g_L d_{Wq} d^c  \bar  \sigma_{\mu \nu}   u W_{\mu\nu}^-  + \hc   \right ]  
\nnl && 
- {h \over 4 v^2} \left [ 
\sum_{f \in u,d}   \tilde d_{Gf}  f^c  \bar  \sigma_{\mu \nu} T^a   f   \widetilde G_{\mu\nu}^a 
+ e  \sum_{f \in u,d,e}   \tilde d_{Af}  f^c  \bar  \sigma_{\mu \nu}   f   \widetilde A_{\mu\nu} 
\right . \nnl && \left . 
+ \sqrt{g_L^2 + g_Y^2}\sum_{f \in u,d,e}   \tilde d_{Zf}    f^c  \bar  \sigma_{\mu \nu}    f  \widetilde Z_{\mu\nu}
+ \sqrt {2} g_L  \tilde d_{Wq} d^c  \bar  \sigma_{\mu \nu}   u  \widetilde W_{\mu\nu}^- + \hc    \right ]. 
\eea 
For Higgs decays into four light fermions, the dipole-like contributions do not interfere with the SM amplitudes due to the different helicity structure. 
Therefore, corrections to the decay width enter quadratically in $d_{Vf}$, and should be neglected.  
Furthermore, as a consequence of the linearly realized electroweak symmetry in the  $D=6$ Lagrangian, the parameters $d_{Vf}$ are proportional to the respective dipole moments which  are stringently constrained by experiment, especially for light fermions. 
For these two reasons, it is safe to neglect the dipole-like Higgs interactions for the sake of LHC analyses.

Finally, $D=6$ operators produce several more interactions involving the single  Higgs boson field, for example Higgs couplings to 3 gauge bosons. 
Observable effects of these couplings are extremely suppressed, therefore they are not listed here. 
Moreover, new interactions involving two (or three) Higgs boson fields appear in the Lagrangian, 
and they are relevant for an EFT description of double Higgs production \cite{Contino:2012xk,Dolan:2012rv,McCullough:2013rea,Englert:2014uqa,Goertz:2014qta,Azatov:2015oxa,Grober:2015cwa}. 
This review is focused on single Higgs production, therefore  multi-Higgs couplings are not listed;  see Ref.~\cite{HXSWGbasis} for the relevant expressions in the Higgs basis.

I close  this section with a brief discussion of the validity range of this approach.  
Formally, EFT is an expansion in powers of the scale $\Lambda$ suppressing higher-dimensional operators. 
Since the independent couplings in \eref{ind}  arise from $D=6$ operators, they  are formally of order $v^2/\Lambda^2$. 
The rule of thumb is that the EFT approach to Higgs physics is valid if $\Lambda \gtrsim v$, which translates to  $|c_i|  \lesssim 1$ and $\delta y_f \lesssim v/m_f$ for the independent couplings. 
However,  a detailed analysis of this issue is much more tricky and depends on the kinematic region probed by a given observable. 
For example, for observables probing the high $\sqrt{s}$ or high $p_T$ tail of differential distributions the validity range will be different than for inclusive observables. 
See Ref.~\cite{Biekoetter:2014jwa} for a more in-depth discussion of these issues.
In this review I restrict to the Higgs signal strength observables in various production modes, which are typically dominated by $\sqrt{s} \sim m_h$. 
Moreover, I am dodging the question of the validity range because it is assumed from the onset that higher-dimensional operators provide small corrections on top of SM contributions. 
Consequently, I will only take into account  corrections to the observables that are linear in the parameters in \eref{ind}, 
which corresponds to retaining only $\cO(\Lambda^{-2})$ effects in the EFT expansion.\footnote{%
Typically, $\cO(\Lambda^{-4})$ effects should be neglected in the context of   $D=6$ effective Lagrangian, as  they may receive contributions from $D=8$ operators.
The exception is  the observables where the SM contribution is suppressed or vanishes, in which case $D=6$ operators contribute  at  $\cO(\Lambda^{-4})$, while contributions of higher-order operators are suppressed by more powers of $\Lambda$. 
One example is the lepton-flavor violating Higgs decays into 2 fermions where the SM contribution is exactly zero.  
In this review I focus on the observables where the SM contribution is dominant.  
}  
Incidentally, the LHC so far confirms that the SM is a decent first approximation of the Higgs sector, and deviations due to new physics are small.  


\section{Observables}
\setcounter{equation}{0} 
\label{sec:obs}

Consider the Higgs boson produced at the LHC via the process $X$, and subsequently decaying to the final state $Y$. 
It is possible, to an extent, to isolate  experimentally different Higgs boson production modes and decays channels.  
The LHC collaborations typically quote the Higgs  signal strength relative to the SM one in a given channel, here denoted as $\mu_{X;Y}$.  
Thanks to the narrow width of the Higgs boson, the production and decay can be separated:\footnote{%
Except in off-shell Higgs processes \cite{Caola:2013yja}. 
However, given the current precision, these processes do not impose any meaningful constraints within the EFT framework \cite{Englert:2014aca,Cacciapaglia:2014rla,Azatov:2014jga,Englert:2014ffa}. 
}
\beq
\mu_{X;Y} = 
{\sigma(pp \to X) \over \sigma (pp \to X)_{\rm SM}  } 
{\Gamma(h \to Y) \over \Gamma(h \to Y)_{\rm SM}} 
{  \Gamma (h \to {\rm all})_{\rm SM}  \over   \Gamma (h \to {\rm all}) } .
\eeq

Below I  summarize how the Higgs production and decays depend on the parameters in the effective Lagrangian.  
These  formulas  allow one to derive experimental constraints on the EFT parameters. 
This kind of approach to LHC Higgs data was pioneered in Refs.~\cite{Carmi:2012yp,Azatov:2012bz} and perfected in Refs.~\cite{Espinosa:2012ir,Rauch:2012wa,Giardino:2012ww,Ellis:2012rx,Azatov:2012rd,Farina:2012ea,Klute:2012pu,Corbett:2012dm,Giardino:2012dp,Ellis:2012hz,Montull:2012ik,Espinosa:2012im,Carmi:2012in,Banerjee:2012xc,Bonnet:2012nm,Plehn:2012iz,Djouadi:2012rh,Batell:2012ca,Corbett:2012ja,Choudhury:2012tk,Belanger:2012gc,Cheung:2013kla,Falkowski:2013dza,Giardino:2013bma,Ellis:2013lra,Djouadi:2013qya,Dumont:2013wma,Bechtle:2013xfa,Belanger:2013xza,Ciuchini:2013pca,Artoisenet:2013puc,Heinemeyer:2013tqa,Chpoi:2013wga,Pomarol:2013zra,Belusca-Maito:2014dpa,Baak:2014ora,Ellis:2014jta,Flament:2015wra}. 
As discussed at the end of \sref{effl}, only linear corrections in the independent couplings are kept,  while quadratic corrections are ignored. 
For this reason only CP-even couplings appear in these formulas (the CP-odd ones enter inclusive observables  only at the quadratic level). 
Moreover, I only include $D=6$ corrections at the {\em tree level} and I ignore new physics effects suppressed by a loop factor.
The exception is the gluon fusion production process which is computed at the next-to-leading order in the $D=6$ parameters. 
Unless  noted otherwise, I give the inclusive production and decay rates. 
Note that the  signal strength quoted by experiments may depend on  analysis-specific cuts, which may slightly  change the dependence on the effective theory parameters.

\bc \bf Production \ec 

For the relevant partonic processes of Higgs production at the LHC, the cross section relative to the SM one depends on the effective theory parameters as follows: 
\bi
\item Gluon fusion (ggh), $g g \to h $:
\beq
{\sigma_{ggh} \over \sigma_{ggh}^{\rm SM}} \simeq \left | 1 + {\hat c_{gg} \over  c_{gg}^{\rm SM} } \right |^2 , 
\eeq 
where 
\bea 
\label{eq:cgghat}
\hat c_{gg} &\simeq&   
c_{gg} +  {1 \over 12 \pi^2}  \left [  \delta y_u  A_f \left (m_h^2 \over 4 m_t^2 \right ) + \delta y_d A_f \left (m_h^2 \over 4 m_b^2 \right ) \right ] ,
\nnl
c_{gg}^{\rm SM} &\simeq&   
{1 \over 12 \pi^2}  \left [ A_f \left (m_h^2 \over 4 m_t^2 \right ) + A_f \left (m_h^2 \over 4 m_b^2 \right ) \right ] ,
 \nnl 
A_f(\tau) &\equiv& \frac{3}{2\tau^2} \left [  (\tau-1)f(\tau)  + \tau \right ], 
\nnl
f(\tau) &\equiv&  \left\{ \begin{array}{lll} 
{\rm arcsin}^2\sqrt{\tau} && \tau \le 1 \\ -\frac{1}{4}\left[\log\frac{1+\sqrt{1-\tau^{-1}}}{1-\sqrt{1-\tau^{-1}}}-i\pi\right]^2 && \tau > 1 \end{array}\right. .
\eea 
As discussed in Ref.~\cite{Gori:2013mia}, in this case it is appropriate to calculate  $c_{gg}^{\rm SM}$ at the leading order in QCD because then the large k-factors,  approximately common for $c_{gg}$ and $\delta y_u$, cancel in the ratio.\footnote{%
Accidentally, with the SM parameters used in this review,  the dependence on $\delta y_d$ is also captured with a decent accuracy by this procedure.   
One can compare \eref{ggh} to  NLO QCD results in Ref.~\cite{Harlander:2013qxa}, where the coefficient in front of  $\delta y_d$ is found to be $-0.06$ for $\sqrt{s} = 8$~TeV,  and $-0.05$ for $\sqrt{s} = 14$~TeV. 
}  
Numerically,   
\beq
\hat c_{gg}  \simeq   c_{gg} + \left ( 8.7 \delta y_u  - (0.3 - 0.3i) \delta y_d \right ) \times 10^{-3}, 
\qquad c_{gg}^{\rm SM} \simeq  (8.4+0.3 i) \times 10^{-3}, 
\eeq  
\beq
\label{eq:ggh}
{\sigma_{ggh} \over \sigma_{ggh}^{\rm SM}} \simeq 1  +  237 c_{gg}  + 2.06 \delta y_u - 0.06  \delta y_d. 
\eeq 
\item Vector boson fusion (VBF),  $q q \to h qq$: 
\label{eq:vbf}
\bea
{\sigma_{VBF} \over \sigma_{VBF}^{\rm SM}} &\simeq & 
1 + 1.49 \delta c_w + 0.51 \delta c_z   -  \bvec 1.08 \\ 1.11  \\ 1.23  \evec c_{w\Box}  - 0.10  c_{ww}  - \bvec 0.35 \\ 0.35 \\  0.40 \evec   c_{z \Box} 
 \nnl & &  
-0.04 c_{zz}  -0.10 c_{\gamma \Box} -  0.02  c_{z\gamma}
 \nnl & \to &  
 1 +  2 \delta c_z  - 2.25  c_{z\Box}  - 0.83  c_{zz} +  0.30 c_{z\gamma} +   0.12 c_{\gamma \gamma} .
 \eea 
The numbers in the columns  multiplying $c_{w\Box}$ and $c_{z\Box}$ refer to the LHC collision energy of $\sqrt{s} =$7, 8,~and~13~TeV; for other parameters the dependence is weaker. 
The expression after the arrow arises due to replacing the dependent couplings by the independent ones in \eref{ind}. 
Each LHC Higgs analysis uses somewhat different cuts to isolate the VBF signal, and the relative cross section slightly depends on these cuts.   
The result in \eref{vbf}  has been computed  numerically by simulating the parton-level process in  MadGraph5 \cite{Alwall:2014hca} at the tree level  with the cuts  $p_{T, q} > 20~$GeV, $|\eta_q| < 5$ and  $m_{qq} > 250$~GeV. 
Replacing the last cut by  $m_{qq} > 500$~GeV affects the numbers at the level of $5 \%$. 
\item Vector boson associated production (Vh), $q \bar q \to Vh$, where $V=W,Z$,
\bea
\label{eq:vhrates}
{\sigma_{W h} \over \sigma_{Wh}^{\rm SM}} & \simeq & 
1 + 2 \delta c_w  + \bvec  6.39 \\ 6.51  \\ 6.96  \evec c_{w\Box}   +  \bvec 1.49\\ 1.49 \\ 1.50 \evec c_{ww}  
\nnl  & \to  & 
1 + 2 \delta c_z 
+ \bvec  9.26 \\ 9.43 \\ 10.08 \evec c_{z\Box}   +  \bvec 4.35 \\ 4.41 \\ 4.63 \evec c_{zz}  -  \bvec 0.81 \\ 0.84 \\ 0.93 \evec c_{z \gamma}  -  \bvec 0.43 \\ 0.44 \\ 0.48 \evec c_{\gamma \gamma}
\nnl 
{\sigma_{Z h} \over \sigma_{Z h}^{\rm SM}} & \simeq & 
1 + 2 \delta c_z 
+ \bvec  5.30 \\ 5.40 \\ 5.72 \evec c_{z\Box}   +  \bvec 1.79 \\ 1.80 \\ 1.82 \evec c_{zz}  +  \bvec 0.80 \\ 0.82 \\ 0.87 \evec c_{\gamma \Box} +  \bvec 0.22 \\ 0.22 \\ 0.22 \evec c_{z \gamma},
\nnl  & \to  & 
1 + 2 \delta c_z 
+ \bvec  7.61 \\ 7.77 \\ 8.24 \evec c_{z\Box}   +  \bvec 3.31 \\ 3.35 \\ 3.47 \evec c_{zz}  -  \bvec 0.58 \\ 0.60 \\ 0.65 \evec c_{z \gamma}  +  \bvec 0.27 \\ 0.28 \\ 0.30 \evec c_{\gamma \gamma}.
\nnl 
\eea 
The numbers in the columns refer to the LHC collision energy of $\sqrt{s} =$7, 8,~and~13~TeV. 
\item Top  pair associated production, $g g \to h t \bar t$:
\beq
{\sigma_{tth} \over \sigma_{tth}^{\rm SM}} \simeq  1 + 2 \delta y_u.  
\eeq
\ei

\bc \bf Decay \ec 

\bi
\item $\bf h \to f \bar f$. 
Higgs boson decays  into 2 fermions occur at the tree level in the SM  via the Yukawa couplings in \eref{SMh}. 
In the presence of  $D=6$ operators they are  affected via the corrections to the Yukawa couplings in \eref{hff}:
\beq
{\Gamma_{cc} \over \Gamma_{cc}^{\rm SM}}   \simeq 1 + 2 \delta y_u, 
\qquad 
{ \Gamma_{bb}\over \Gamma_{bb}^{\rm SM} }  \simeq 1 + 2 \delta y_d,  
\qquad 
{\Gamma_{\tau \tau} \over \Gamma_{\tau \tau}^{\rm SM}}   \simeq   1 + 2 \delta y_e,  
\eeq 
where I abbreviate $\Gamma(h \to Y) \equiv \Gamma_Y$. 

\item $\bf h \to V V$. 
In the SM, Higgs decays into on-shell gauge bosons:  gluon pairs $gg$, photon pairs $\gamma \gamma$, and $Z \gamma$ occur only at the one-loop level.    
In the presence of $D=6$ operators these decays are corrected already at the tree level  by the 2-derivative contact interactions of the Higgs boson with two vector bosons in \eref{hvv}.  
The relative decay widths are given by 
\beq
\label{eq:gammavv}
{ \Gamma_{VV} \over  \Gamma_{VV}^{\rm SM}}  \simeq  \left |1 +  {\hat c_{vv} \over c_{vv}^{\rm SM}}  \right |^2, 
\qquad    
vv  \in  \{ gg, \gamma \gamma, z \gamma \} ,  
\eeq 
where 
\bea 
\label{eq:cvvsm}
\hat c_{\gamma \gamma} &= &  c_{\gamma \gamma}, \qquad  c_{\gamma \gamma}^{\rm SM} \simeq -8.3  \times 10^{-2},  
\nnl 
\hat c_{z \gamma} &=&  c_{z \gamma}, \quad c_{z \gamma}^{\rm SM} \simeq -5.9  \times 10^{-2}, 
\eea 
while $\hat c_{gg}$ and $c_{gg}^{\rm SM}$ are defined in \eref{cgghat}. 
Note that  contributions to  $\Gamma_{\gamma \gamma}$ and $\Gamma_{z \gamma}$ arising due to corrections to the SM Higgs couplings to the W bosons and fermions are not included in \eref{cvvsm}, unlike in \eref{cgghat}.
The reason is that, for these processes,  corrections from $D=6$ operators are included at the tree level only. 
If these particular one-loop corrections  were included, one should also consistently include {\em all} one-loop corrections to this process arising at the $D=6$ level, some of which are divergent and require renormalization. 
The net result would be to redefine $\hat c_{\gamma \gamma} =  c_{\gamma \gamma}^{\rm ren.}  -  0.11   \delta c_w  + 0.02 \delta y_u + \dots $, and  $\hat c_{z \gamma} =  c_{z \gamma}^{\rm ren.}  -0.06 \delta c_w + 0.003 \delta y_t + \dots$.
Here  "ren." stands for ``renormalized" and the dots stand for a dependence on other Lagrangian parameters ($c_{ww}$, $c_{w\Box}$, and corrections to triple gauge couplings). 
A full next-to-leading order computation of these processes have not been yet attempted in the literature.

\item $\bf h \to 4 f$. 
The decay process $h\to 2 \ell 2 \nu$ (where $\ell$  here stands for charged leptons) proceeds via intermediate W bosons. 
 The relative width is given by 
\bea 
{\Gamma_{2 \ell 2\nu} \over  \Gamma_{2\ell 2\nu}^{\rm SM}} 
&\simeq & 
1 + 2 \delta c_w  + 0.46 c_{w \Box}   - 0.15 c_{ww}  
\nnl 
&\to &  
1 + 2 \delta c_z   + 0.67 c_{z\Box} + 0.05 c_{zz}  - 0.17 c_{z\gamma} -  0.05 c_{\gamma\gamma}.  
\eea  

In the SM, the decay process $h \to 4 \ell$ proceeds at the tree-level via intermediate Z  bosons. 
In the presence $D=6$ operators,  intermediate photon contributions may also arise at the tree level.
If that is the case, the decay width diverges due to the photon pole. 
Below I quote the relative  width $\bar \Gamma(h \to 4 \ell)$ regulated by imposing  the cut  $m_{\ell \ell} > 12$~GeV  on the invariant mass of same-flavor lepton pairs: 
\bea
\hspace{-1cm} {\bar \Gamma_{4 \ell} \over  \bar \Gamma_{4 \ell}^{\rm SM}} 
&\simeq & 
1 + 2 \delta c_z   +  \bvec 0.41 \\ 0.39 \evec c_{z\Box}  - \bvec 0.15 \\ 0.14  \evec c_{zz}   + \bvec 0.07 \\ 0.05 \evec  c_{z \gamma}  - \bvec  0.02 \\ 0.02 \evec  c_{\gamma \Box}  + \bvec < 0.01 \\  0.03 \evec c_{\gamma \gamma}
\nnl  &\to & 
1+ 2 \delta c_z +  \bvec 0.35 \\  0.32 \evec  c_{z \Box} -  \bvec 0.19 \\ 0.19 \evec  c_{zz} +  \bvec  0.09 \\ 0.08  \evec  c_{z\gamma} +  \bvec 0.01 \\ 0.02 \evec c_{\gamma\gamma} .
\eea 
The numbers in the columns correspond to the $2 e 2 \mu$ and $4 e/\mu$ final states, respectively. 
The difference between these two is numerically irrelevant  in the total width, but may be important for differential distributions, especially  regarding the $c_{\gamma \gamma}$ dependence  \cite{Chen:2015iha}.  
The dependence on the $m_{\ell \ell}$ cut is weak; very similar numbers are obtained if  $m_{\ell \ell} > 4$~GeV is imposed instead. 
\ei

Given the partial widths,  the branching fractions can be computed as ${\rm Br}_Y= \Gamma_Y/\Gamma(h \to {\rm all})$, where the total decay width  is given by 
\beq
{\Gamma(h \to {\rm all}) \over \Gamma(h \to {\rm all}) } \simeq 
 {\Gamma_{bb}\over \Gamma_{bb}^{\rm SM} }  {\rm Br}_{bb}^{\rm SM} 
 +   {\Gamma_{cc}\over \Gamma_{cc}^{\rm SM} }  {\rm Br}_{cc}^{\rm SM}
 +  {\Gamma_{\tau \tau}\over \Gamma_{\tau \tau}^{\rm SM} }  {\rm Br}_{\tau \tau}^{\rm SM}  
+  {\Gamma_{WW^*}\over \Gamma_{WW^*}^{\rm SM} }  {\rm Br}_{WW^*}^{\rm SM} 
+  {\Gamma_{ZZ^*}\over \Gamma_{ZZ^*}^{\rm SM} }  {\rm Br}_{ZZ^*}^{\rm SM}
+  {\Gamma_{gg}\over \Gamma_{gg}^{\rm SM} }  {\rm Br}_{gg}^{\rm SM}.  
\eeq  
Note that, in line with the basic assumption of no new light particles,  
there is no additional  contributions to the Higgs width other than from the SM decay channels.  
In particular, the invisible Higgs width is absent in this EFT framework (except for the small SM contribution arising via $h \to Z Z^* \to 4 \nu$).

\section{Current Constraints}
\setcounter{equation}{0} 
\label{sec:constraints}

In this section I present the constraints on the independent couplings characterizing the Higgs boson couplings in the dimension six EFT Lagrangian. 
A disclaimer is in order. 
The objective is to illustrate what is the constraining power of the present data. 
As we will see, the existing  data is not yet good enough to even constrain all the couplings inside the EFT validity range. 
In the future, as the measurements become more precise and more information is available,  this kind of analysis will become fully consistent. 

\subsection{Data}
\label{sec:data}

I first review the experimental data used to constrain the effective theory parameters. 
In the best of all worlds, the LHC collaborations would quote a multi-dimensional likelihood function for the signal strength $\mu_{X;Y}$ for all production modes and decay channels, separately for each LHC collision energy. 
This would allow one to  consistently use available experimental information, including non-trivial  correlations between the different $\mu$'s. 
Although the manner in which  the LHC data is presented has been constantly improving, we are not yet in the ideal world. 
For these reasons, constraining Higgs couplings  from existing data  involves inevitably somewhat arbitrary assumptions and approximations.   
Nevertheless, thanks to the fact that the experimental uncertainties are still statistics-dominated in most cases, one should expect that these approximations do not affect the results in a dramatic way.

The measurements of the Higgs signal strength $\mu$  included in  this analysis are summarized in \tref{exp}.
They are separated according to the final state (channel) and the production mode. 
For the all-inclusive production mode (total)  I use the value of  $\mu$ quoted in the table.\footnote{%
CMS does not quote the best-fit $\mu$ in the $Z \gamma$ channel. 
The value in \tref{exp} was obtained by digitizing the plot showing the expected and observed 95\%~CL limits on $\mu$ in function of $m_h$, extracting  the values at $m_h = 125$~GeV, and using these to calculate the best-fit $\mu$ assuming the uncertainties are Gaussian.   
This is a dire reminder of how Higgs fits had to be done back in the early 2010s. 
 }
 The same is true for  $\mu$ in a specific production mode (Wh, Zh, tth), in which case I ignore correlations with other production channels. 
In the remaining  cases  $\mu$ is quoted for illustration only, and more information is included in the analysis. 
2D stands for two-dimensional likelihood functions in the plane $\mu_{\rm ggh+tth}$-$\mu_{\rm VBF+Vh}$.   
Since, the contribution of the $Vh$ production mode is subleading with respect to the VBF one, I combine the separate measurements of the $Vh$ signal strength (whenever it is given) with the 2D likelihood,  ignoring the correlation between the two.  
For the diphoton final state I construct five-dimensional likelihood function in the space spanned by $(\mu_{ggh}, \mu_{tth}, \mu_{\rm VBF}, \mu_{Wh},\mu_{Zh})$ using the signal strength in all diphoton event categories (cats.), using the known contribution of each production mode to each  category. 
In many channels there is a certain degree of arbitrariness as to  which set of results (inclusive,1D, 2D, or cats.) to include in the fit; 
here the strategy  is to choose the set that maximizes the available information about various EFT couplings.

\begin{table}[t]
\begin{center}
\renewcommand*{\arraystretch}{1.2} 
\begin{tabular}{|c|l|c|c|}
\hline
\multicolumn{4}{|c|}{\bf ATLAS} \\ \hline 
Channel  & $\mu$ & Production & Ref. 
\\ \hline 
$\gamma \gamma$ & $1.17^{+0.28}_{-0.26}$ & cats.   &  \cite{Aad:2014eha}
\\ \hline 
$Z \gamma$ & $2.7^{+4.6}_{-4.5}$ & total   &  \cite{atlascoup}
\\ \hline 
$Z Z^*$ & $1.46^{+0.40}_{-0.34}$ & 2D   &  \cite{Aad:2014eva}
\\ \hline 
$W W^*$ & $1.18^{+0.24}_{-0.21}$ & 2D   &  \cite{ATLAS:2014aga}
\\ \cline{2-4} 
& $2.1^{+1.9}_{-1.6}$ & Wh   &  \cite{atlaswwvh}
\\ \cline{2-4} 
& $5.1^{+4.3}_{-3.1}$ & Zh   &  \cite{atlaswwvh}
\\ \hline 
$\tau \tau$ & $1.44^{+0.42}_{-0.37}$ & 2D   &  \cite{Aad:2015vsa}
\\ \hline 
$b b$  & $1.11^{+0.65}_{-0.61}$ & Wh   &  \cite{Aad:2014xzb}
\\ \cline{2-4} 
  & $0.05^{+0.52}_{-0.49}$ & Zh  &  \cite{Aad:2014xzb}
\\ \cline{2-4} 
  & $1.5^{+1.1}_{-1.1}$ & tth   &  \cite{Aad:2015gra}
\\ \hline 
 $\mu \mu$  & $-0.7^{+3.7}_{-3.7}$ & total   &  \cite{atlascoup}
\\ \hline 
 multi-$\ell$  & $2.1^{+1.4}_{-1.2}$ & tth   &  \cite{atlasml}
\\ \hline 
\end{tabular}
\quad 
\begin{tabular}{|c|l|c|c|}
\hline
\multicolumn{4}{|c|}{\bf CMS } \\ \hline 
Channel  & $ \mu$ & Production & Ref. 
\\ \hline 
$\gamma \gamma$ & $1.12^{+0.25}_{-0.22}$ & cats.   &  \cite{Khachatryan:2014ira}
\\ \hline 
$Z \gamma$ & $-0.2^{+4.9}_{-4.9}$ & total   &  \cite{Chatrchyan:2013vaa}
\\ \hline 
$Z Z^*$ & $1.00^{+0.29}_{-0.29}$ & 2D   &  \cite{Khachatryan:2014jba}
\\ \hline 
$W W^*$ & $0.83^{+0.21}_{-0.21}$ & 2D   & \cite{Khachatryan:2014jba}
\\ \cline{2-4} 
& $0.80^{+1.09}_{-0.93}$ & Vh  & \cite{Khachatryan:2014jba}
\\ \hline 
$\tau \tau$ & $0.91^{+0.28}_{-0.28}$ & 2D   &  \cite{Khachatryan:2014jba}
\\ \cline{2-4} 
& $0.87^{+1.00}_{-0.88}$ & Vh   &  \cite{Khachatryan:2014jba}
\\ \cline{2-4} 
 & $-1.3^{+6.3}_{-5.5}$ &  tth   &     \cite{Khachatryan:2014qaa}
\\ \hline 
$b b$  & $0.89^{+0.47}_{-0.44}$   & Vh   &  \cite{Khachatryan:2014jba}
\\ \cline{2-4} 
  & $1.2^{+1.6}_{-1.5}$ & tth  &  \cite{Khachatryan:2015ila}
\\ \hline 
 $\mu \mu$  & $0.8^{+3.5}_{-3.4}$ & total   &  \cite{Khachatryan:2014aep}
\\ \hline 
 multi-$\ell$  & $3.8^{+1.4}_{-1.4}$ & tth  &  \cite{Khachatryan:2014qaa}
\\ \hline 
\end{tabular}
\end{center}
\caption{
\label{tab:exp}
The LHC Higgs results used in the fit. See \sref{data} for explanations.}
\end{table}

\subsection{Fit}

Using the dependence of the signal strength  on EFT parameters worked out in \sref{obs} and the LHC data in \tref{exp} one can constrain all {\em CP-even} independent Higgs couplings  in \eref{ind}.\footnote{%
To constrain the CP-odd couplings $\sin \phi_f$ and $\tilde c_{vv}$ within the EFT framework  one should study the differential distributions in multi-body Higgs decays where these couplings enter at the linear level \cite{Bishara:2013vya,Chen:2014pia,Dolan:2014upa,Chen:2014ona,Chen:2014gka,Beneke:2014sba,Demartin:2014fia,Berge:2014sra}.
}
In the Gaussian approximation near the best fit point I find the following constraints: 
\beq
\label{eq:fit}
\bvec \delta c_z \\  c_{zz} \\  c_{z\Box} \\  c_{\gamma \gamma} \\ c_{z\gamma} \\  c_{gg} \\  \delta y_u  \\ \delta y_d \\  \delta y_e   \evec  = 
\bvec  -0.12 \pm 0.20 \\ 0.5 \pm  1.8 \\ -0.21 \pm 0.82 \\  0.014 \pm 0.029  \\  0.01 \pm 0.10  \\ -0.0056 \pm 0.0028 \\ 0.55 \pm 0.30 \\ -0.42 \pm 0.45  \\ -0.17 \pm 0.35  \evec, 
\eeq 
where the uncertainties correspond to $1 \sigma$.  
The correlation matrix is 
\beq
\label{eq:rho}
\rho = \left ( \begin{array}{ccccccccc}
 1. & -0.23 & 0.17  & -0.62 & -0.18 & 0.16 & 0.09 & 0.88 & 0.63  \\
\cdot&1. & -0.997 & 0.85  & 0.23 &  0.13 & 0.17 & -0.47 & -0.81  \\
\cdot&\cdot&1. & -0.82 & -0.21 & -0.15 & -0.17 & 0.41 & 0.78   \\
\cdot&\cdot&\cdot&1. & 0.27 & 0.02  & 0.09 & -0.79 & -0.92   \\
\cdot &\cdot&\cdot&\cdot&1. & 0.01 & 0.02 & -0.22 & -0.26  \\
\cdot&\cdot&\cdot&\cdot&\cdot&1. & -0.81 & 0.21 & 0.03   \\
\cdot&\cdot&\cdot&\cdot&\cdot&\cdot&1. & 0.05 & -0.06  \\
\cdot&\cdot&\cdot&\cdot&\cdot&\cdot&\cdot&1. & 0.82  \\
\cdot&\cdot&\cdot&\cdot&\cdot&\cdot&\cdot&\cdot&1.
\end{array} \right ).
\eeq
Using the above central values  $c_0$,  uncertainties $\delta c$,   and the correlation matrix $\rho$,  
one can reconstruct the 9-dimensional likelihood function near the best fit point: 
\beq
\label{eq:chi2}
\chi^2 \simeq   \sum_{ij}[c -  c_0]_{i} \sigma^{-2}_{ij}   [c - c_0]_{j}, \qquad \sigma^{-2}_{ij} \equiv  [[\delta c]_i \rho_{ij} [ \delta c]_j]^{-1}. 
\eeq

\subsection{Discussion}

As  one can see from \eref{fit},  certain EFT parameters are very weakly constrained by experiment, with order one deviations from the SM being allowed. 
In other words, the current data cannot even constrain all the parameters to be within the EFT validity range.
This violates the initial assumption that the $D=6$ operators give a small correction on top of the SM. 
For this reason,  the results in \eref{fit} should not be taken at face value. 
In particular, one should conclude that there is currently no model-independent constraints at all on $c_{zz}$ and $c_{z \Box}$.
Indeed, including corrections to observables that are quadratic in these parameters would completely change the central values and the uncertainties. 
This signals a sensitivity of the fit to operators with $D>6$.  
Furthermore, the experimental  constraints in the $Z \gamma$ channel are still too weak to justify the linear approximation. 
Again, including quadratic EFT corrections would significantly affect the constraints on $c_{z \gamma}$.
To a lesser extent, the sensitivity to higher-order EFT corrections is also true for deformations along  the $\delta y_f$ directions. 

Nevertheless, the results in \eref{fit} are of some value. 
First of all, they demonstrate that {\em certain} EFT parameters are strongly constrained. 
This is true especially for $c_{gg}$ and $c_{\gamma \gamma}$ who are constrained at the $10^{-3}$ level. 
Next, the fit  in \eref{fit} identifies ``blind" directions in the space of the EFT parameters that are weakly constrained by current data. 
The most dramatic example is the approximate degeneracy  along the line $c_{zz} \approx -2.3 c_{z \Box}$, as witnessed by the $\approx 1$ entry in the correlation matrix in \eref{rho}. 
More data is needed to lift this degeneracy. 
To this end, extremely helpful pieces of information can be extracted from differential distributions in $h\to 4 \ell$ decays \cite{Stolarski:2012ps,Chen:2012jy,Chen:2013ejz,Chen:2014pia,Gonzalez-Alonso:2014eva,Gonzalez-Alonso:2015bha}, 
as well as in the Vh \cite{Ellis:2013ywa,Godbole:2013saa,Englert:2013vua,Isidori:2013cga,Biekoetter:2014jwa,Godbole:2014cfa,Ellis:2014dva} and VBF \cite{Djouadi:2013yb,Maltoni:2013sma,Edezhath:2015lga} production.  
A consistent, model-independent EFT approach to Higgs differential distributions has not yet been implemented in LHC analyses, but the CMS collaboration made first steps in this direction\cite{Khachatryan:2014kca}. 
Note also the large correlations between $\delta y_d$ and other parameters.
This happens because $\delta y_d$ strongly affects the total Higgs width (via the $h \to b \bar b$ partial width) and this way it affects the signal strength in all Higgs  decay channels. 
More precise measurements of  the signal strength in the $h \to b \bar b$ channel should soon alleviate this degeneracy. 
Finally, there is the strong correlation between $c_{gg}$ and $\delta y_t$ which has been extensively discussed in the Higgs fits literature. 
In the future, that degeneracy will be lifted by better measurements of the tth signal strength, and by measurements of the Higgs $p_T$ distribution in the gluon fusion production mode \cite{Azatov:2013xha,Grojean:2013nya,Buschmann:2014twa,Dawson:2014ora,Dawson:2015gka} (see also \cite{Arnesen:2008fb} for an earlier work in this direction). 

Finally, the importance of the fit is in the fact that  the likelihood in \eref{chi2} can be combined with other datasets that constrain the same EFT parameters. 
In this case, one may obtain stronger bounds that will push the parameters into the EFT validity range. 
For example, one can use constraints on cubic self-couplings of electroweak gauge bosons \cite{Corbett:2013pja,Pomarol:2013zra,Masso:2014xra,Ellis:2014jta,Falkowski:2014tna,Bobeth:2015zqa,Gonzalez-Alonso:2015bha}. 
These are customarily parametrized by  3 parameters $\delta g_{1,z}$, $\delta \kappa_\gamma$, $\lambda_z$ \cite{Hagiwara:1986vm} which characterize deviations of these self-couplings from the SM predictions.
Now, in the EFT Lagrangian with $D=6$ operators the first two parameters are related to the Higgs couplings. 
In the Higgs basis one finds \cite{HXSWGbasis}: 
\bea
 \delta  g_{1,z} &=& 
{1 \over 2 (g_L^2 - g_Y^2)} \left [   c_{\gamma\gamma} e^2 g_Y^2 + c_{z \gamma} (g_L^2 - g_Y^2) g_Y^2  - c_{zz} (g_L^2 + g_Y^2) g_Y^2  - c_{z\Box} (g_L^2 + g_Y^2) g_L^2 \right ], 
 \nnl
 \delta \kappa_\gamma  &=& - {g_L^2 \over 2} \left ( c_{\gamma\gamma} {e^2 \over g_L^2 + g_Y^2}   + c_{z\gamma}{g_L^2  - g_Y^2 \over g_L^2 + g_Y^2} - c_{zz} \right ).
\eea  
Therefore, model-independent constraints on triple gauge couplings imply additional constraint on the EFT parameters characterizing the Higgs couplings. 
In particular, Ref.~\cite{Falkowski:2014tna} argues that, after marginalizing over $\lambda_z$,  the single and pair W boson production in LEP-2 implies the bounds  $\delta g_{1,z} = -0.83 \pm 0.34$, $\delta \kappa_\gamma = 0.14 \pm 0.05$ with the correlation coefficient $[\rho]_{\delta g_{1,z}}^{\delta \kappa_\gamma} = -0.71$. 
Combining this bound with the likelihood in \eref{chi2}  the degeneracy between $c_{zz}$ and $c_{z \Box}$ is lifted, and  
one obtains much stronger bounds: $c_{zz} = 0.22 \pm  0.18$, $c_{z\Box} = -0.08 \pm 0.09$, $[\rho]_{c_{zz}}^{c_{z \Box}} = -0.76$. 
More constraints of this type, for example model-independent constraints on triple gauge couplings from the LHC, could further improve the limits on Higgs couplings within the EFT approach.
As soon as  more precise Higgs and di-boson data from the 13 TeV LHC run start arriving, it should be possible to constrain all the 9 parameters in \eref{fit} safely within the EFT validity range.

\section{Closing Words}
\setcounter{equation}{0} 
\label{sec:con}

The Higgs boson has been discovered, and for the remainder of this century we will study its properties. 
Precision measurements of Higgs couplings and determination of their tensor structure is an important part of the physics program at the LHC and future colliders.
Given that no slightest hint for a particular scenario beyond the SM has emerged so far,  it is important to (also) perform these studies in a model-independent framework. 
The EFT approach described here, with the SM extended by dimension six operators,  provides a perfect tool to this end. 
 
One should be aware that Higgs precision measurements cannot probe new physics at very high scales. 
For example, LHC Higgs measurements are sensitive to new physics at $\Lambda \sim 1$~TeV at the most. 
This is not too  impressive, especially  compared to the new physics reach of flavor observables or even electroweak precision tests. 
However, Higgs physics  probes a subset of operators that are often not accessible by other searches. 
For example, for most of the 9 parameters in \eref{fit} the {\em only} experimental constraints come from Higgs physics. 
It is certainly conceivable that new physics talks to the SM via the Higgs portal, and it will first manifest itself within this particular class of $D=6$ operators.
If this is the case, we must not miss it.

\section*{Acknowledgements}
 
I am supported by the ERC Advanced Grant Higgs@LHC.


\bibliographystyle{JHEP}
\bibliography{indiansummer}

\end{document}